\documentclass[twocolumn, reprint, superscriptaddress, amsmath,amssymb,aps,prl]{revtex4-2}

\usepackage{graphicx}
\usepackage{dcolumn}
\usepackage{bm}
\usepackage[hidelinks,colorlinks=true,linkcolor=blue,citecolor=black]{hyperref}
\usepackage[mathlines]{lineno}

\usepackage{braket}
\usepackage{color}
\usepackage{float}
\usepackage{amsmath}
\usepackage[dvipsnames]{xcolor}

\begin{document}

\title{Universal Manipulation of Quantum Synchronization in Spin Oscillator Networks}

\affiliation{Department of Physics, Renmin University of China, Beijing, 100872, China}
\affiliation{Shenzhen Key Laboratory of Ultraintense Laser and Advanced Material Technology, Center for Intense Laser Application Technology, and College of Engineering Physics, Shenzhen Technology University, Shenzhen 518118, China}
\affiliation{Computational Materials Science Research Team, RIKEN Center for Computational Science (R-CCS), Kobe, Hyogo 650-0047, Japan}
\affiliation{Quantum Science Center of Guangdong-Hongkong-Macao Greater Bay Area (Guangdong), Shenzhen 518045, China}

\author{Shuo Dai}
\affiliation{Department of Physics, Renmin University of China, Beijing, 100872, China}
\affiliation{Shenzhen Key Laboratory of Ultraintense Laser and Advanced Material Technology, Center for Intense Laser Application Technology, and College of Engineering Physics, Shenzhen Technology University, Shenzhen 518118, China}

\author{Zeqing Wang}
\affiliation{Computational Materials Science Research Team, RIKEN Center for Computational Science (R-CCS), Kobe, Hyogo 650-0047, Japan}
\affiliation{Quantum Science Center of Guangdong-Hongkong-Macao Greater Bay Area (Guangdong), Shenzhen 518045, China}

\author{Liang-Liang Wan}
\affiliation{Shenzhen Key Laboratory of Ultraintense Laser and Advanced Material Technology, Center for Intense Laser Application Technology, and College of Engineering Physics, Shenzhen Technology University, Shenzhen 518118, China}

\author{Weidong Li}
\affiliation{Shenzhen Key Laboratory of Ultraintense Laser and Advanced Material Technology, Center for Intense Laser Application Technology, and College of Engineering Physics, Shenzhen Technology University, Shenzhen 518118, China}

\author{Augusto Smerzi}
\affiliation{Shenzhen Key Laboratory of Ultraintense Laser and Advanced Material Technology, Center for Intense Laser Application Technology, and College of Engineering Physics, Shenzhen Technology University, Shenzhen 518118, China}
\affiliation{QSTAR and INO-CNR and LENS, Largo Enrico Fermi 2, 50125 Firenz, Italy}
 
\author{Ran Qi}
\email{Corresponding author: qiran@ruc.edu.cn}
\affiliation{Department of Physics, Renmin University of China, Beijing, 100872, China}
 
\author{Jianwen Jie}
\email{Corresponding author: Jianwen.Jie1990@gmail.com}
\affiliation{Shenzhen Key Laboratory of Ultraintense Laser and Advanced Material Technology, Center for Intense Laser Application Technology, and College of Engineering Physics, Shenzhen Technology University, Shenzhen 518118, China}
\date{\today}

\begin{abstract}
Quantum synchronization (QS) in open many-body systems offers a promising route for controlling collective quantum dynamics, yet existing manipulation schemes often rely on dissipation engineering, which distorts limit cycles, lacks scalability, and is strongly system-dependent. Here, we propose a universal and scalable method for continuously tuning QS—from maximal synchronization under isotropic interactions to complete synchronization blockade (QSB) under fully anisotropic coupling in spin oscillator networks. Our approach preserves intrinsic limit cycles and applies to both few-body and macroscopic systems. We analytically show that QS arises solely from spin flip-flop processes and their higher-order correlations, while anisotropic interactions induce non-synchronizing coherence. A geometric QS measure reveals a macroscopic QSB effect in the thermodynamic limit. The proposed mechanism is experimentally feasible using XYZ interactions and optical pumping, and provides a general framework for programmable synchronization control in complex quantum networks and dynamical phases of matter. 
\end{abstract}

\maketitle
{\em Introduction.---} Synchronization, a ubiquitous phenomenon in nature, plays a crucial role in understanding dynamics across fields ranging from science and engineering to societal systems \cite{pikovsky2001synchronization,strogatz2024nonlinear,Kuramoto1984ChemicalOW}. Recent quantum advances in engineered dissipations \cite{Harrington2022,PRL2017QZE,PRL2018reset,Yang2025,PRL2024Liu,PRB2023Wang} have expanded the study of synchronization to open quantum systems \cite{PRL2013VdP,PRL2014VdP,PRL2019VdP,PRR2020VdP,PRL2013OM1,PRE2024Sudler,PRL2018VdP,PRA2022squzzing,PRA2018coldatom,PRA2019optomechanical,PRA2015QED,Weiss_2016,PRE2012OM,PRA2014OM,PRL2012NOexp,Science2007NOexp,li2025ion,PRL2017kerr,PRA2018QSB,PRL2018Spin1,PRL2018QN,PRL2023MQSE,Solanki2023PRA,PRA2024Tobia,PRA2025QSBandNR,PRA2025QSandQFI,PRA2020two,PRB2024No_Go,PRB2009QED,PRL2020exp,Zhang2023PRR,PRR2020SPin1,PRE2020heat,Tan2022halfintegervs,PRR2020hybrid,kumar2025,Zhu2024,Xia2025,li2025twobody,PRA2023Li,PRL2022noise,PRL2024energy,PRL2024measure,PRL2024Lie}. Key developments include the demonstration of hallmark features of synchronization, such as limit cycle (LC) and Arnold tongue, in continuous-variable nonlinear quantum oscillators \cite{PRL2013VdP,PRL2014VdP,PRL2019VdP,PRR2020VdP,PRL2017kerr,PRA2019optomechanical,PRE2024Sudler,PRL2013OM1,PRA2018QSB,PRL2018VdP,PRA2018coldatom,PRA2015QED,Weiss_2016,PRE2012OM,PRA2014OM,PRL2012NOexp,Science2007NOexp,li2025ion,PRA2022squzzing}, discrete spin systems \cite{PRL2018Spin1,PRL2018QN,PRL2023MQSE,Solanki2023PRA,PRA2024Tobia,PRA2025QSBandNR,PRA2025QSandQFI,PRR2020SPin1,Zhang2023PRR,PRB2009QED,PRL2020exp,PRA2020two,PRB2024No_Go,Tan2022halfintegervs,PRE2020heat} and hybrid boson-spin systems \cite{PRR2020hybrid}, as well as their realizations in experiments involving cold atoms \cite{PRL2020exp}, trapped ions \cite{Zhang2023PRR,li2025ion}, IBM-Q \cite{PRR2020SPin1}, optomechanical platforms \cite{PRA2019optomechanical} and nuclear spins \cite{PRA2022nuclear}. The quantum generalization of classical synchronization measures and phase space theory \cite{PRL2013VdP,PRL2018Spin1,PRL2024Lie} suggests that quantum synchronization (QS) fundamentally represents a new form of quantum coherence and correlations \cite{galve2017quantum,PRA2013Spin}. This understanding has broadened the applications of QS in quantum thermodynamics and phase transitions \cite{book2018,PRB2010Spin,PRE2020heat}, while also forging intriguing connections to quantum entanglement \cite{PRA2015QED,galve2017quantum,he_entanglement_2024,PRR2020hybrid}, quantum metrology \cite{PRA2025QSandQFI,shen_fisher_2023}, topology \cite{wachtler2023topological,wachtler2024topological,mr1f-v8cv} and time crystals \cite{nadolny_nonreciprocal_2025,solanki_exotic_2024}.

One particularly counterintuitive discovery among these advances is the suppression of synchronization between identical quantum oscillators—a striking departure from the robust phase-locking observed in their classical counterparts~\cite{PRL2017kerr}. This phenomenon, known as quantum synchronization blockade (QSB)~\cite{PRL2017kerr,PRA2018QSB,PRL2018Spin1}, is a uniquely quantum effect that eludes any semiclassical description, bearing conceptual similarities to the Pauli blockade~\cite{PRL2022Tong}, photon blockade~\cite{photonblockadePRL1997}, and Rydberg blockade~\cite{PRL2010entanglment}. Although QSB has been observed in a wide range of systems~\cite{PRL2017kerr,PRL2018Spin1,PRL2018QN,PRA2018QSB,PRL2023MQSE,Solanki2023PRA,PRA2024Tobia,PRA2025QSBandNR,PRA2025QSandQFI,PRA2020two,PRB2024No_Go}, the physical conditions required for its emergence vary considerably. For instance, QSB in a single driven-spin oscillator requires specific dissipation symmetry \cite{Solanki2023PRA,PRL2018Spin1,PRA2020two}, while QSB between spin-1 oscillators necessitates balanced dissipation \cite{PRL2018QN}, QSB between qubit oscillators requires identical dissipation ratios \cite{PRB2024No_Go}, and QSB between Kerr-anharmonic oscillators demands frequency resonance \cite{PRL2017kerr}. 

Despite these diverse realizations, a unifying feature is the crucial role of symmetry—particularly dissipative symmetry—in suppressing phase locking, typically through quantum interference and rotational invariance. Nevertheless, a general theoretical framework and a universal control strategy—applicable across diverse platforms and capable of continuously tuning synchronization from its maximal value to complete QSB—remain elusive. Developing such a mechanism is crucial for extending QS to complex and heterogeneous quantum networks~\cite{ARENAS200893,Zhu_2015,PRL2023MQSE,nadolny_nonreciprocal_2025,Lorenzo_2022,LI2017121,Antonelli_2017,Lohe_2010,Manzano2013,Tan2022halfintegervs}, thereby enabling scalable and programmable control of collective synchronization dynamics. Here, we propose a universal approach to QS based on interaction control. As illustrated in Fig.~\ref{fig1}, tuning spin interactions from fully isotropic to fully anisotropic continuously drives the system from maximal synchronization to complete QSB. Our analysis reveals that the core mechanism underlying QS—namely, phase locking—arises exclusively from spin flip-flop processes and their higher-order correlations, whereas the anisotropic components induce coherence unrelated to phase locking. 

This insight also clarifies why general quantum correlation measures, such as entanglement and discord, are inadequate indicators of QS. We rigorously demonstrate these findings in both few-body and macroscopic spin networks and further extend the mechanism to oscillators with arbitrary spin. Finally, we discuss potential experimental implementations using Rydberg atoms and other platforms featuring XYZ-type interactions and tunable gain–loss dissipation. Our results fill a critical gap in the field by establishing a unifying and scalable strategy for the programmable control of QS.

\begin{figure}[t]
\centering
\includegraphics[width=8.6cm]{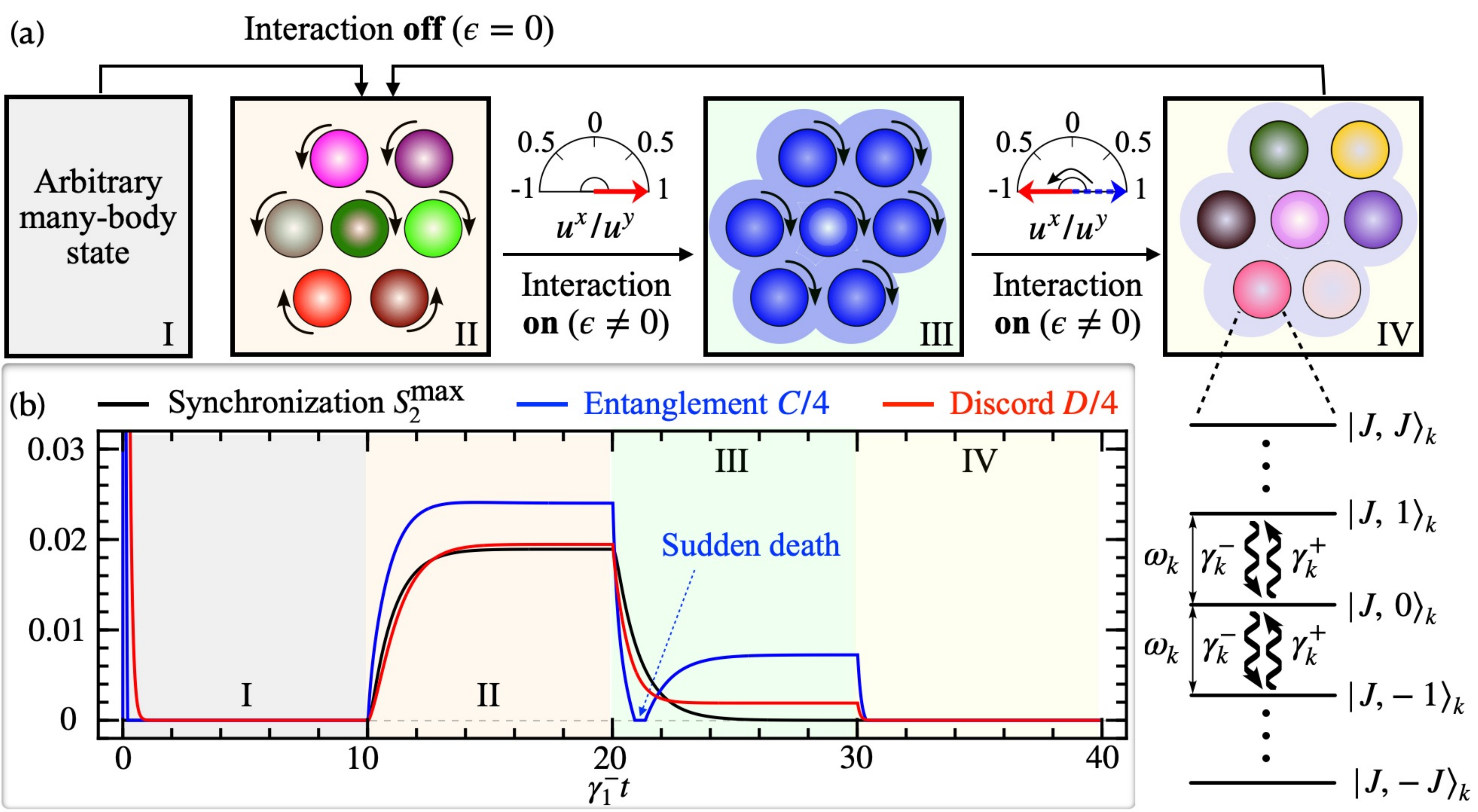}
\caption{\label{fig1} {\color{black} (a) Illustration of the universal interaction-based control of spin networks, each spin is subject to both damping ($\gamma^{-}_{j}$) and gain ($\gamma^{+}_{j}$) dissipations. Without interaction (I, IV), each spin independently evolves into a limit cycle state (II). When interactions are isotropic ($u^{x}/u^{y}=1$), the system reaches maximum synchronization (III); when anisotropic ($u^{x}/u^{y}=-1$), synchronization is blocked (IV). (b) Two spin-1 oscillator example. Synchronization $S_{2}^{\text{max}}$, entanglement $C$ and discord $D$, exhibits similar behavior across stages I, II, and IV. However, at stage III, neither $C$ and $D$ displays the quantum blockade effect as $S_{2}^{\text{max}}$ does. Notably, $C$ experiences a sudden-death followed by a revival to a nonzero value. Parameters: $\gamma^{+}_{2}=\gamma^{-}_{1},\gamma^{+}_{1}=\gamma^{-}_{2}=100\gamma^{-}_{1},\epsilon=0.1\gamma^{-}_{1},u^{z}=0,\Delta_{12}=\omega_{1}-\omega_{2}=0$}}
\end{figure}

{\em System and quantum synchronization.---} We consider a network of $N$ dissipative spin-$J$ systems, coupled via the Heisenberg XYZ interaction, resulting in the Hamiltonian ($\hbar=1$)
\begin{eqnarray}\label{eqHamiltonian} 
\hat{H} = \sum_{k=1}^{N} \omega_{k} \hat{J}_{k}^{z} + \sum_{k<l}^{N} \sum_{\alpha=x,y,z} \epsilon u_{k,l}^{\alpha} \hat{J}_{k}^{\alpha} \hat{J}_{l}^{\alpha}.
\end{eqnarray} 
Here, $\omega_{k}$ and $\hat{J}_{k}^{\alpha}$ are the natural frequency and spin operators for the $k$th spin, with $\epsilon$ representing the overall interaction strength and $u_{k,l}^{\alpha} \in [-1, 1]$. Each spin experiences independent dissipations, consisting of gain (+) and damping (-), described by $\mathcal{L}_{k}^{(\pm)}\hat{\rho}=\gamma^{\pm}_{k} \mathcal{D}[\hat{O}^{\pm}_{k}]\hat{\rho}/2$ with the Lindblad super-operator $\mathcal{D}[\mathcal{\hat{A}}]\hat{\rho} = \mathcal{\hat{A}}\hat{\rho} \mathcal{\hat{A}}^\dagger - \{ \mathcal{\hat{A}}^\dagger \mathcal{\hat{A}}, \hat{\rho} \}/2$. Consequently, the system's evolution is governed by the Lindblad master equation

\begin{eqnarray}\label{eqLME} 
\frac{d\hat{\rho}}{dt} = -i[\hat{H}, \hat{\rho}]+\mathcal{L}\hat{\rho}= -i[\hat{H}, \hat{\rho}] + \sum_{k=1}^{N} (\mathcal{L}_{k}^{(+)}+\mathcal{L}_{k}^{(-)})\hat{\rho}.~~~~
\end{eqnarray} 

Synchronization encompasses a broad range of research topics \cite{pikovsky2001synchronization}. In this work, we focus primarily on QS based on LCs \cite{PRL2018Spin1,PRL2018QN,PRL2023MQSE,Solanki2023PRA,PRA2024Tobia,PRA2025QSBandNR,PRA2025QSandQFI,PRR2020SPin1,Zhang2023PRR,PRB2009QED,PRL2020exp,PRA2020two,PRB2024No_Go}, which is analogous to classical chaos synchronization \cite{BOCCALETTI20021,PRL1996CS,PRE1996EC}. The $Q$-function, 
\begin{eqnarray}\label{Qfunction} 
Q(\vec{\phi},\vec{\theta},\hat{\rho})= \left(\frac{2J+1}{4\pi}\right)^{N} \bra{\vec{\phi},\vec{\theta}}\hat{\rho}\ket{\vec{\phi},\vec{\theta}}, 
\end{eqnarray} 
provides a phase portrait of the quantum state $\hat{\rho}$ in the Husimi $Q$ representation \cite{PRL2018Spin1,PRL2018QN}. $\ket{\vec{\phi},\vec{\theta}}\equiv\bigotimes_{k=1}^{N}\ket{\phi_{k},\theta_{k}}$ represents the product coherent spin state with the most classical states $\ket{\theta,\phi}=\exp(-i\phi\hat{J}^{z})\exp(-i\theta\hat{J}^{y})\ket{J,J}$.  As shown in Fig. \ref{fig1}(a), when the interaction is turned off, the system decorrelates to the product steady state $\hat{\rho}_{ss}^{(0)}=\otimes_{k=1}^{N}\hat{\rho}_{ss,k}^{(0)}$ of Eq. (\ref{eqLME}). $\hat{\rho}^{(0)}_{ss,k}$ is verified as a valid LC because its $Q$-function is independent of $\phi_{k}$, indicating that $\phi_{k}$ is a free phase \cite{PRL2018Spin1}. In other words, regardless of the initial many-body state (I), each spin stabilizes to its respective LC steady state and periodically accumulates the free phase as a self-sustained oscillator (II).

QS can be viewed as the process of adjusting free phases to achieve phase-locking between them. The $S$-function is proposed as a QS measure \cite{PRA2024Tobia}, obtained by tracing out the energy information encoded in phase $\theta$,
\begin{eqnarray}\label{Srel}
    S_N(\vec{\phi})=(\prod_{j=1}^{N}\int_{0}^{\pi}d\theta_{j}\sin\theta_{j})Q(\vec{\phi},\vec{\theta},\hat{\rho})-\frac{1}{(2\pi)^{N}}.
\end{eqnarray}
This function represents the distribution of the free phase $\vec{\phi}=(\phi_{1},\phi_{2},\cdots,\phi_{N})$, where the position and height of the distribution's peak respectively quantify the locked phase $\vec{\phi}^{(0)}$ and the QS  strength, $S^{\text{max}}_{N}= \max_{\vec{\phi}}S_N(\vec{\phi}) = S_N(\vec{\phi}^{(0)})$. An inhomogeneous distribution of $Q$ and $S$ with respect to the free phase $\vec{\phi}$ produces a nonvanishing $S^{\text{max}}_{N}$, resulting in a synchronized state. Conversely, the significant suppression of $S^{\text{max}}_{N}$ indicates QSB. 

Our main result is the proposal of a universal method for manipulating QS via interaction control. Specifically, as illustrated in Fig.~\ref{fig1}(a), maximal QS (III) and QSB (IV) in spin oscillator networks can be achieved by tuning the interactions to be fully isotropic ($u^{x}/u^{y}=1$) and fully anisotropic ($u^{x}/u^{y}=-1$), respectively. To illustrate this method, we next focus on spin-1 oscillators, which have been verified as the smallest systems capable of forming limit cycles with phase-symmetric pure steady states that are completely immune to dissipative noise \cite{PRL2018Spin1,PRL2018QN}. Nevertheless, the results we present are generic, and we will later generalize them to systems with arbitrary spin (Supplementary Materials \cite{footnote_SM}).

\begin{figure}
\centering
\includegraphics[width=8.6cm]{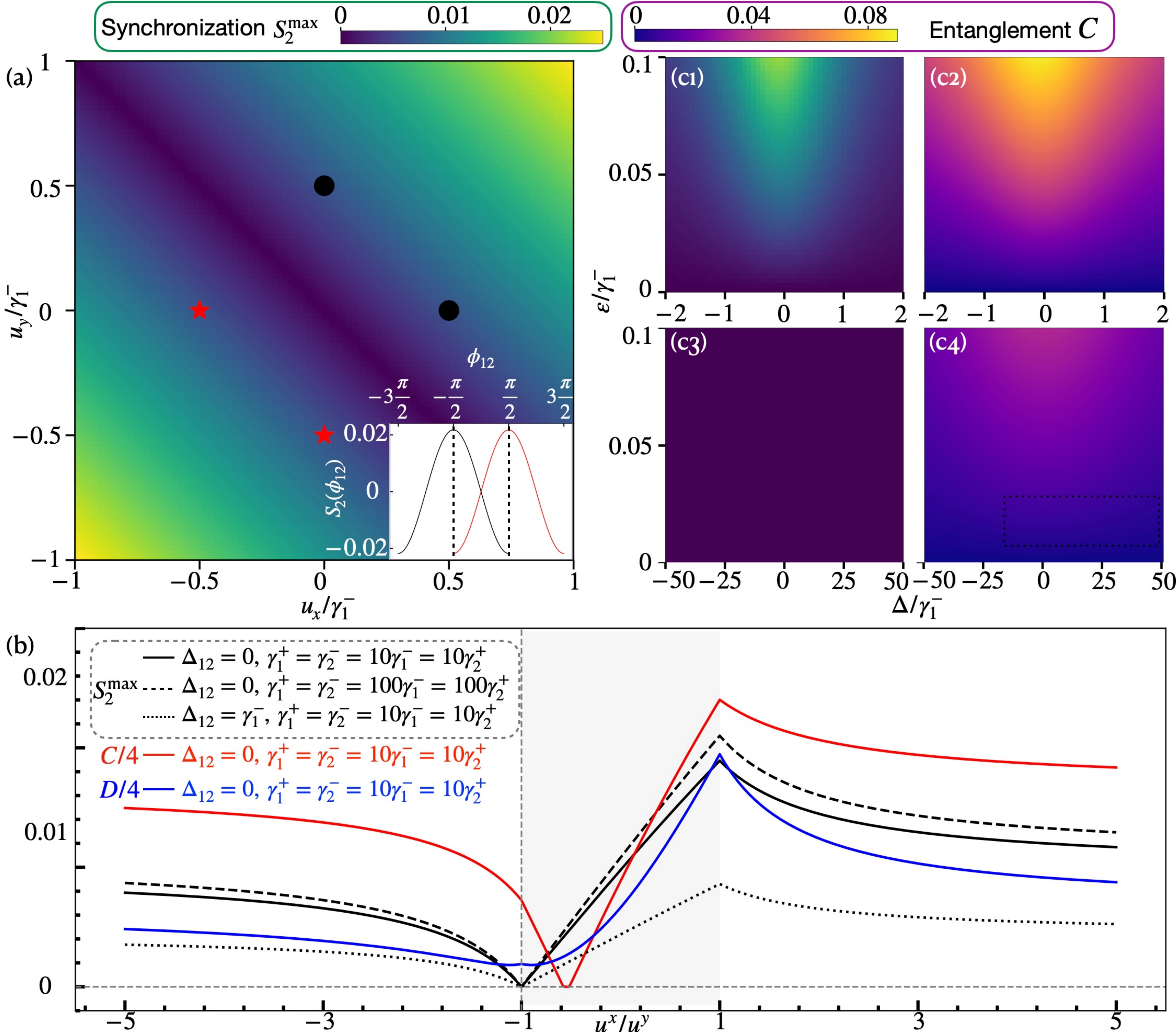}
\caption{\label{fig2} (a) Synchronization measure $S^{\text{max}}_{2}$ for two resonant spin-1 oscillators with imbalanced gain and damping. Synchronization is suppressed when $u^{x}/u^{y} = -1$, indicating a blockade. Insets show phase locking at $-\pi/2$ and $\pi/2$, corresponding to repulsive (black lines, circles) and attractive (red lines, stars) couplings, respectively.
(b) Control of synchronization $S^{\text{max}}_{2}$, entanglement $C$, and mutual information $I$ by tuning the anisotropy ratio $u^{x}/u^{y}$. The shaded region highlights the regime where only synchronization can be linearly tuned from full blockade to its maximum.
(c) Arnold tongues for synchronization (c1, c3) and entanglement (c2, c4) under $u^{x}/u^{y} \neq -1$ (c1-c2) and $u^{x}/u^{y} = -1$ (c3-c4). Synchronization vanishes under purely anisotropic interaction (c3), whereas entanglement persists with a visible tongue structure(c4).
Parameters: (a-c) $u^z = 0$; (a-b) $\epsilon = 0.1\gamma^{-}_{1}$; (a,c) $\gamma^{+}_{1} = \gamma^{-}_{2} = 100\gamma^{-}_{1} = 100\gamma^{+}_{2}$; (b) $|u^{x}+u^{y}|+|u^{x}-u^{y}|=1$.}
\end{figure}

{\em Two spin-1 oscillators.---} This case corresponds to setting $N=2$ and $J=1$ in Eq. (\ref{eqLME}), with $\hat{O}_{j}^{\pm}=\hat{J}_{j}^{\pm}\hat{J}_{j}^{z}$ taken as the jump operators \cite{PRL2018Spin1}. To focus the discussion on the relative phase  $\phi_{12}=\phi_{1}-\phi_{2}$, we further define $S_{2}(\phi_{12})=\int_{0}^{2\pi}d\phi_{2}S_{2}(\phi_{12}+\phi_{2},\phi_{2})$. Figure \ref{fig2}(a) shows the QS measure $S^{\text{max}}_{2}$ for resonant oscillators ($ \omega_{1} = \omega_2 $) with imbalanced gain and damping strengths ($ \gamma_{1,2}^{-} \neq\gamma_{1,2}^{+}  $) \cite{PRL2018QN}. The symmetry pattern shows that $S^{\text{max}}_{2}$ remains invariant under the transformation $(u^{x}, u^{y}) \leftrightarrow (-u^{x}, -u^{y})$, while the relative free phase locks at $\pi/2$ ($-\pi/2$) when $u^{x} + u^{y}$ is positive (negative) (see the inset figure). The two anti-diagonal corners, corresponding to larger $|u^{x}+u^{y}|$, reveal that stronger couplings result in stronger synchronization, consistent with the Arnold tongue \cite{PRL2018Spin1,PRL2018QN}. In contrast, QS is significantly suppressed when $u^{x}$ and $u^{y}$ have opposite signs, and vanishes entirely when the interaction is purely anisotropic. This finding is particularly noteworthy, as the similar squeezing configurations in van der Pol oscillator systems were previously found to be highly favorable for observing QS \cite{PRL2018VdP,PRA2022squzzing}.

To further investigate the effects of anisotropic interaction, Fig. \ref{fig2}(b) shows results obtained by fixing the overall interaction strength $\epsilon$ while varying the degree of anisotropy in the interaction. We find that, regardless of detuning and dissipation, QSB consistently occurs when the interaction is purely anisotropic, and $S^{\text{max}}_{2}$ reaches its maximum when the interaction is purely isotropic. Additionally, the QS measure $S^{\text{max}}_{2}$ varies linearly within the range $u^{x}/u^{y} \in [-1, 1]$. Outside this range, QS nonlinearly saturates at a value. These findings indicate that $u^{x}/u^{y}$ is an effective tuning parameter for linearly controlling QS.

To understand the QSB mechanism in this resonant case, we perform a perturbation expansion in the overall interaction strength $\epsilon$, as synchronization is primarily a perturbative response to coupling \cite{pikovsky2001synchronization}, and derive the following analytical expression:
\begin{eqnarray}\label{xxx}
 S_{2}(\phi_{12})\approx\frac{9\pi\epsilon(u^x+u^y)}{256}\left(\frac{1}{\gamma_1^d+\gamma_2^g}-\frac{1}{\gamma_1^g+\gamma_2^d}\right)\sin\phi_{12},
\end{eqnarray}
which illustrates why purely anisotropic interactions act as a novel trigger for QSB. Furthermore, this result aligns with \cite{PRL2018QN}, which showed that QS is suppressed due to destructive interference when all spin oscillators have identical dissipation differences.

To understand the QSB mechanism for general cases, we derive Eq. (\ref{Srel}) for two spins, yielding:  
\begin{eqnarray}\label{Stwospin1}
 S_{2}(\phi_{12})&=&\frac{9\pi}{256}\left|\left\langle \hat{J}_1^+\hat{J}_2^-\right\rangle\right|\cos(\phi_{12}-\phi')\nonumber\\
 &&+\frac{1}{16\pi}\left|\left\langle (\hat{J}_1^{+}\hat{J}_2^{-})^{2}\right\rangle\right|\cos(2\phi_{12}-\phi''),~~
\end{eqnarray}
where $\phi'=\arg[\langle\hat{J}^{+}_{1}\hat{J}^{-}_{2}\rangle]$ and $\phi''=\arg[\langle(\hat{J}^{+}_{1}\hat{J}^{-}_{2})^{2}\rangle]$. Equation (\ref{Stwospin1}) demonstrates that only spin flip-flop correlations contribute to QS, whether at first or higher orders. This emphasizes that QS is intrinsically tied to the locking of relative phases, where a phase difference accumulates only when one spin flips while the other flops during their evolution. In contrast, if both spins flip up or down simultaneously, they accumulate a phase sum, which does not contribute to phase locking. Rewriting the interaction in Eq. (\ref{eqHamiltonian}) as $\epsilon[(u^{x} + u^{y})(\hat{J}_{1}^{+} \hat{J}_{2}^{-} + \text{h.c.}) + (u^{x} - u^{y})(\hat{J}_{1}^{+} \hat{J}_{2}^{+} + \text{h.c.})]/4$ clearly reveals that the isotropic component corresponds to spin flip-flop processes, which conserve total magnetization and thus contribute to QS. In contrast, the anisotropic component induces spin-non-conserving (flip-flip and flop-flop) processes, i.e., spin fluctuations, which break magnetization conservation and are therefore ineffective for QS. 

The fact that QS is a form of quantum correlation, as shown in Eq. (\ref{Stwospin1}), is consistent with the finding that QS information cannot be extracted from the reduced density matrix of each individual spin \cite{PRL2018QN}. This motivates the exploration of other quantum correlations, such as quantum discord ($D$) \cite{PRLQD2001} and concurrence ($C$), a measure of entanglement \cite{PRL1997QC}, to determine whether they can qualify as QS measures. Figure \ref{fig1}(b) and \ref{fig2}(b) show that when $u^{x}/u^{y}$ is close to $-1$, $C$ and $D$ not only do not vanish but also exhibit nonlinear behaviors that qualitatively differ from $S_{2}^{\text{max}}$. In contrast, all correlations show no qualitative differences as $u^{x}/u^{y}$ approaches 1. These features are consistent with Fig. \ref{fig2}(c), where both $S_{2}^{\text{max}}$ and $C$ exhibit a tongue structure when $u^{x} \neq -u^{y}$. However, when $u^{x} = -u^{y}$, the entanglement tongue displayed by $C$ becomes entirely unrelated to synchronization. These results indicate that \( C \) and \( D \) capture broader quantum correlations than \( S_{2}^{\mathrm{max}} \), and can only serve as reliable measures of QS when the interaction is nearly isotropic. This strongly supports that the \( S \)-function provides a particularly effective and selective measure of QS in spin systems compared to other available indicators.

\begin{figure}
    \centering
    \includegraphics[width=8.5cm]{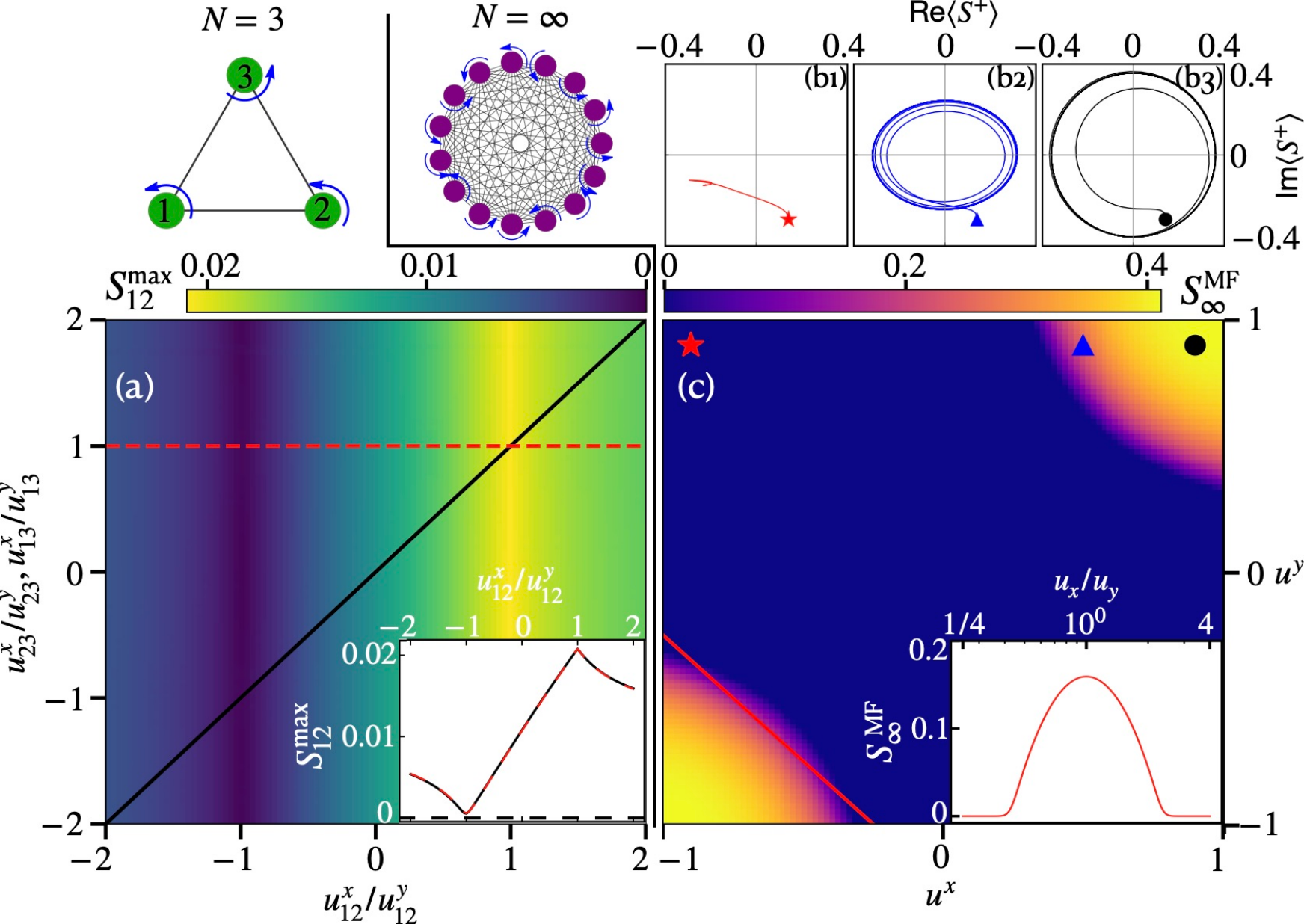}
    \caption{\label{fig3} Synchronization in spin-1 networks for \( N = 3 \) (a) and \( N=\infty \) (b,c). (a) Synchronization measure \( S^{\mathrm{max}}_{12} \) between 1th and 2nd spins. %1 and spin 2 as a function of interaction anisotropy. 
    The inset compares the universal control scenario along the red dashed line and the black diagonal, showing that the synchronization behavior is insensitive to the presence of spin 3. (b) Dynamics of a large-scale network in the thermodynamic limit for three representative cases: fully anisotropic (b1), partially isotropic (b2), and fully isotropic (b3). Symbols indicate the initial states, which correspond to the representative cases shown in (c). (c) Synchronization measure \( S_\infty^{\mathrm{MF}} \), defined as the area enclosed by the limit-cycle trajectory. The inset (c1) corresponds to the red solid line and illustrates how increasing anisotropy suppresses synchronization even when the total isotropic strength is held fixed. Parameters: (a) $\epsilon = 0.1\gamma^{-}_{1}$, $|u_{1,2,3}^{x}+u_{1,2,3}^{y}|+|u_{1,2,3}^{x}-u_{1,2,3}^{y}|=1$,$\gamma^{+}_{1} = \gamma^{-}_{2} = 10\gamma^{+}_{3} = 5\gamma^{-}_{3} = 100\gamma^{-}_{1} = 100\gamma^{+}_{2}$, $\Delta_{12}=\Delta_{23}=0$; (b) $\epsilon = 6\gamma^{+}$, $\gamma^{-}= 10\gamma^{+}$.}
\end{figure}

{\em Spin-1 oscillator network.---} A central question for spin oscillator networks is whether their complex topological connectivity undermines the QSB effect. All-to-all quantum networks offer a foundational framework for studying collective phenomena in quantum systems by utilizing all available interactions and are well aligned with current experimental implementations \cite{Li2024,Miller2024}. 

We first analyze an all-to-all network of three spin-1 oscillators by fully solving Eq.~(\ref{eqLME}), as shown in Fig.~\ref{fig3}(a). It shows that the QS between spins 1 and 2 is nearly unaffected by the indirect couplings \( u_{13} \) and \( u_{23} \), and is instead solely controlled by the anisotropy ratio of their direct interaction, \( u_{12}^{x} / u_{12}^{y} \) (see inset). This control behavior closely resembles that in Fig.~2(b), except that when \( u_{12}^{x}/u_{12}^{y} = -1 \), a small but finite residual QS persists (see the horizontal dashed line in the inset). This incomplete QSB arises from higher-order interaction processes mediated by the third spin, which weakly break the effective two-spin isolation. We have also verified that such processes become negligible in spin chain networks.

Furthermore, we consider a large-scale all-to-all network compromises identical spin-1 oscillators as shown in Fig. \ref{fig3}(b–c), where the exponential growth of the Hilbert space renders full quantum simulations intractable. Thus we following \cite{PRL2023MQSE}, employ a mean-field treatment in the thermodynamic limit (\( N = \infty \)), which corresponds to neglecting all operator correlations and applying a product-state ansatz. Under this approximation, the dynamics of each identical spin in the rotating frame is governed by a nonlinear master equation $ d\hat{\rho}/dt = -i[\epsilon(u_x \hat{J}^x \langle \hat{J}^x\rangle + u_y \hat{J}^y \langle \hat{J}^y\rangle), \hat{\rho}] + (\gamma^+ \mathcal{D}[\hat{J}^+ \hat{J}^z] + \gamma^- \mathcal{D}[\hat{J}^- \hat{J}^z])\hat{\rho}$ \cite{footnote_SM}. In the absence of interaction, each spin exhibits a random phase due to intrinsic quantum noise, leading to destructive interference in the collective amplitude. As a result, the mean spin coherence \( \langle J^+ \rangle =\sum_k \langle \hat{J}^+_k \rangle = \sum_{k}|\langle \hat{J}^{+}_{k} \rangle| e^{i\phi_k}\) vanishes, indicating the absence of QS \cite{footnote_noQS}. 

Figure~\ref{fig3}(b) shows the coherence dynamics for three representative cases, with initial states marked by symbols corresponding to the parameters in Fig.~\ref{fig3}(c). The perfectly circular trajectory originates from the fully isotropic case (b3), corresponding to complete QS. As the interaction anisotropy increases, the closed trajectory becomes squeezed into an elliptical shape in the partially isotropic case (b2), and eventually collapses into a fixed point in the fully anisotropic case (b1), signaling the QSB. The nonzero coherence \( |S^+| \) observed in the fixed point arises from a spin-squeezing effect induced by anisotropic interactions, which is unrelated to phase locking.

We therefore define a geometric measure \( S_\infty^{\mathrm{MF}} \), the area enclosed by the closed trajectory, as a measure of QS. Figure~\ref{fig3}(c) clearly shows that QS emerges only in the presence of sufficiently isotropic interactions. Moreover, when the total isotropic strength \( u^x + u^y \) is held constant (red solid line), as illustrated in inset, increasing the anisotropy progressively suppresses QS. Remarkably, this suppression can lead to complete QSB, even without perfect anisotropy. Such QSB does not occur in systems with a finite number of spins and is a distinctive \textit{macroscopic quantum effect} enabled by the thermodynamic limit, further enriching the landscape of universal QS control.

{\em Arbitrary spin oscillator. ---}
Although the computational exponential wall in many-body studies poses a significant challenge, the above discussions on spin-1 can, in principle, be extended to $N$ spin-$J$, with the $Q$-function and $S$-function already defined in Eqs. (\ref{Qfunction}-\ref{Srel}). For a single spin-$J$, the $S$-function simplifies to $S_{1}(\phi)= \text{Tr}[ \hat{c}^{J}(\phi)\hat{\rho}]-1/2\pi$, where $c^{J}_{n,m}(\phi)=e^{i(n-m)\phi} \Gamma[1+J+(n+m)/2]\Gamma[1+J-(n+m)/2]/2\pi\sqrt{(J+n)!(J-n)!(J+m)!(J-m)!}$. Then  $S$-function in Eq. (\ref{Srel}) can then be generalized to $S_{N}^{n}(\vec{\phi})=\text{Tr}[  \otimes_{j=1}^{n}\hat{c}^{J}(\phi_{j})\rho]-1/(2\pi)^{N}$, which quantifies QS among any $n$ spin-$J$ subsystems within the larger $N$-spin system. Since synchronization is primarily a perturbative response to coupling, the steady state $\hat{\rho}_{ss}$ can be expanded as a power series in $\epsilon$,  $\hat{\rho}=\sum_{j=0}\epsilon^{j}\hat{\rho}^{(j)}$, where the zeroth-order term $\hat{\rho}^{(0)}$ is the aforementioned product LC steady state, determined by free Hamiltonian and the dissipation: $\mathcal{L}\hat{\rho}^{(0)}=i[\sum_{k=1}^{N}\omega_{k}\hat{J}_{k}^{z}, \hat{\rho}^{(0)}] $. For higher orders, the interaction couples adjacent terms, governed by:  $ \mathcal{L}\hat{\rho}^{(j+1)} = i[\sum_{k<l}^{N} \sum_{\alpha=x,y,z} \epsilon u_{k,l}^{\alpha} \hat{J}_{k}^{\alpha} \hat{J}_{l}^{\alpha}, \hat{\rho}^{(j)}]$.  We can rigorously show that the two-body QS measure \( S_N^2 \) receives contributions exclusively from correlation terms of the form \(\left\langle (\hat{J}_1^+ \hat{J}_2^-)^p + \mathrm{h.c.} \right\rangle \), with \( p = 1, 2, \dots, 2J \). As a result, anisotropic interactions do not contribute to QS. The detailed discussion are provided \cite{footnote_SM}.

{\em Conclusion and Discussion. ---}
We have achieved universal manipulation of QS in spin oscillator networks via interaction control. Previous dissipation-based approaches typically suffer from several limitations, including distortion of limit-cycle structures, strong system dependence of control conditions, and poor scalability to large systems. In contrast, our interaction-based method offers three key advantages:  
(1) It operates independently of dissipation, thereby preserving and stabilizing intrinsic limit cycles;  
(2) It enables dynamic tuning of interaction anisotropy while maintaining quantum correlations throughout the control process;  
(3) Its linear control framework simplifies system manipulation, enhances dynamical stability, and facilitates scalable QS implementation in large-scale spin networks.

The realization of our scheme requires two key ingredients: XYZ-type spin interactions and engineered gain–loss dissipation, both of which are already experimentally accessible. XYZ interactions have been demonstrated in various platforms—such as adamantane molecules \cite{Li2024}, Rydberg atoms \cite{PRL2023Arrays,PRL2013pumping}, cavity QED systems \cite{Luo2025}, and polar molecular setups \cite{Miller2024}—using Floquet pulse sequences and related techniques. Meanwhile, gain and loss dissipation can be reliably implemented via optical pumping \cite{Harrington2022}, a well-established experimental tool. As detailed in the Supplemental Material~\cite{footnote_SM}, we provide a concrete implementation scheme using Rydberg atoms as a representative example.

Our work establishes a general and scalable strategy for programmable control of quantum synchronization in diverse many-body quantum systems, offering a promising path toward engineering synchronization-based dynamical phases of matter.

\begin{acknowledgments}
This work was supported by the National Natural Science Foundation of China (Grant No. 12575026, 12022405, 11774426 and 12405023), the National Key Research and Development Program of China (Grant No. 2022YFA1405301 and 2018YFA0306502), the Natural Science Foundation of Top Talent of SZTU (Grant No. GDRC202202 and GDRC202312), and the Guangdong Provincial Quantum Science Strategic Initiative (Grant No. GDZX2305006).
\end{acknowledgments}

\bibliography{UMQS_Refs}

\begin{thebibliography}{87}
\bibitem [{foo({\natexlab{b}})}]{footnote_noQS}%
  \BibitemOpen
  \href@noop {} {\bibfield  {journal} {\bibinfo  {journal} {This behavior is
  also supported by the mean-field analysis: for small coupling strengths, the
  system converges to the steady state \( \rho = |J,0\rangle\langle J,0| \),
  which exhibits no phase preference, yielding \( \langle S^{\pm} \rangle = 0
  \). In contrast, finite coupling tends to align the oscillator phases and
  induce a nonzero collective coherence}\ } ({\natexlab{b}})}\BibitemShut
  {NoStop}%
\end{thebibliography}

%%%%%%%%%%%%%%%%%%%%%%%%%%%%%%%%%%%%%%%%%%%%%%%%%%%%%%
\global\long\def\id{\mathbbm{1}}
\global\long\def\ui{\mathbbm{i}}
\global\long\def\ud{\mathrm{d}}

%%%%%%%%%%%%%%%%%%%%%%%%%%%%%%%%%%%%%%%%%%%%%%%%%%%%%%%%%%%%%%%
\setcounter{equation}{0} \setcounter{figure}{0}
\setcounter{table}{0} %\setcounter{page}{1} \makeatletter
\renewcommand{\theparagraph}{\bf}
\renewcommand{\thefigure}{S\arabic{figure}}
\renewcommand{\theequation}{S\arabic{equation}}

\onecolumngrid
\flushbottom
%%%%%%%%%%%%%%%%%%%%%%%%%%%%%%%%%%%%%%%%%%%%%%%%
\newpage
\maketitle
\title{Supplementary Material:\\Universal Manipulation of Quantum Synchronization in Spin Oscillator networks}

\maketitle
\section{Arbitrary spin oscillators}\label{sec:2}
We have investigated quantum synchronization in spin-1 systems under anisotropic interactions. Our results demonstrate a universal scheme for manipulating quantum synchronization, including the emergence of synchronization blockade under fully anisotropic coupling. To assess the generality of this mechanism, we further examine whether the blockade persists in systems with arbitrary spin-$J$.
\subsection{Analytical derivation}
The evolution of an open quantum system is governed by a master equation that incorporates both coherent dynamics via the Hamiltonian and incoherent processes via dissipation. Here, our primary objective is to elucidate how each of these components influences the quantum synchronization measure \( S \). To this end, we perform a perturbative expansion of the density matrix and systematically analyze the contributions at each order. This approach enables us to disentangle the respective roles of the Hamiltonian and dissipation in shaping the behavior of \( S \). Specifically, we consider a model of coupled spin-$J$ oscillators subject to gain and damping processes. The dynamics of the system are governed by the following Lindblad master equation:
\begin{eqnarray}\label{eqLME} 
\frac{d\hat{\rho}}{dt} = -i[\hat{H}, \hat{\rho}]+\mathcal{L}\hat{\rho}= -i[\hat{H}, \hat{\rho}] + \sum_{k=1}^{N} (\mathcal{L}_{k}^{(+)}+\mathcal{L}_{k}^{(-)})\hat{\rho}.
\end{eqnarray} 
where the Hamiltonian is
\begin{eqnarray}\label{eqHamiltonian} 
\hat{H} =\hat{H}_0+\epsilon\hat{V} = \sum_{k=1}^{N} \omega_{k} \hat{J}_{k}^{z} + \sum_{k<l}^{N} \sum_{\alpha=x,y,z} \epsilon u_{k,l}^{\alpha} \hat{J}_{k}^{\alpha} \hat{J}_{l}^{\alpha}.
\end{eqnarray} 
with $\epsilon$ being the overall coupling strength. Here, $\omega_{k}$ and $\hat{J}_{k}^{\alpha}$ are the natural frequency and spin operators for the $k$th spin, with $\epsilon$ representing the overall interaction strength and $u_{k,l}^{\alpha} \in [-1, 1]$. Each spin experiences independent dissipations, consisting of gain (+) and damping (-), described by $\mathcal{L}_{k}^{(\pm)}\hat{\rho}=\gamma^{\pm}_{k} \mathcal{D}[\hat{O}^{\pm}_{k}]\hat{\rho}/2$ with the Lindblad super-operator $\mathcal{D}[\mathcal{\hat{A}}]\hat{\rho} = \mathcal{\hat{A}}\hat{\rho} \mathcal{\hat{A}}^\dagger - \{ \mathcal{\hat{A}}^\dagger \mathcal{\hat{A}}, \hat{\rho} \}/2$.  Then, we expand the steady-state  of Eq. (\ref{eqLME}) in the powers of $\epsilon$,
\begin{equation}
\hat{\rho}_{ss}  = \sum_{j=0}^{\infty} \epsilon^j \hat{\rho}^{(j)}.
\end{equation}
Substituting $\hat{\rho}_{ss}$ into Eq.~(\ref{eqLME}) and collecting terms according to the order of \( \epsilon \), we can find that the zeroth-order density matrix \( \hat{\rho}^{(0)} \) satisfies the following relations:
\begin{equation}
    \mathcal{L}\hat{\rho}^{(0)}=i[\hat{H}_0, \hat{\rho}^{(0)}],~~ \mathcal{L}\hat{\rho}^{(1)} = i[\hat{V}, \hat{\rho}^{(0)}],
\end{equation}
which subsequently allows us to derive the general relation,
\begin{equation}\label{njiuy}
    \mathcal{L}\hat{\rho}^{(j+1)} = i[\hat{V}, \hat{\rho}^{(j)}].
\end{equation}
From these relations, it follows that the zeroth-order density matrix \( \hat{\rho}^{(0)} \) is determined solely by the free Hamiltonian and the dissipation. As a result, \( \hat{\rho}^{(0)} \) is generally straightforward to obtain, and in many cases admits an analytical solution. In contrast, higher-order corrections \( \hat{\rho}^{(j)} \) depend recursively on the lower-order density matrices and the interaction terms.

For an arbitrary state \( \hat{\rho} \), the density matrix elements \( \rho_{\mathbf{m}, \mathbf{n}} \) can be written as:
\begin{equation}
\rho_{\mathbf{m}, \mathbf{n}} = \langle \mathbf{m} | \hat{\rho} | \mathbf{n} \rangle = \langle m_1, ..., m_N | \rho | n_1, ..., n_N \rangle,
\end{equation}
where  $m_j(n_j)=-J,-J+1,\cdots,J$. 
Then the full density matrix can be decomposed as
\begin{equation}\label{zadasd}
\hat{\rho} = \sum_{\mathbf{m},\mathbf{n}}\rho_{\mathbf{m}, \mathbf{n}}  |\mathbf{m}\rangle \langle \mathbf{n} |.
\end{equation}

Now, we consider the jump operators of dissipations for the $k$th spin as
\begin{equation}
\hat{O}^{\pm}_{k}=\hat{J}_{k}^{\pm} \hat{J}_{k}^z,
\end{equation}
for calculating Eq. (\ref{njiuy}), where we need to consider $\mathcal{L}\hat{\rho}^{(j+1)} $ with the Lindblad super-operator $\mathcal{D}[\mathcal{\hat{A}}]\hat{\rho} = \mathcal{\hat{A}}\hat{\rho} \mathcal{\hat{A}}^\dagger - \{ \mathcal{\hat{A}}^\dagger \mathcal{\hat{A}}, \hat{\rho} \}/2$.  For the first part of Lindblad super-operator $\mathcal{\hat{A}}\hat{\rho} \mathcal{\hat{A}}^\dagger $, we have 
\begin{align}
\hat{O}^{\pm}_{k} |\mathbf{m}\rangle &= c_{m_k} |\mathbf{m} \pm \delta \mathbf{e}_k\rangle, 
\end{align}
where \( \mathbf{e}_k \) denotes the unit vector associated with the \( k \)th spin, \( \delta = 1 \) and \( c_{m_k}= m_{k}\sqrt{(J\mp m_{k})(J\pm m_{k}+1)}\) if \( |m_k \pm \delta| \leq J \); otherwise, \( c_{m_k} = 0 \). Then we have
\begin{equation}
\hat{O}^{\pm}_{k}  |\mathbf{m} \rangle \langle \mathbf{n}  | [\hat{O}^{\pm}_{k}]^\dagger = c_{m_k} c_{n_k} |\mathbf{m} \pm \delta \mathbf{e}_k\rangle \langle \mathbf{n} \pm \delta \mathbf{e}_k |.
\end{equation}
This indicates that the first part of the Lindblad super-operator couples only the matrix elements lying on the same off-diagonal of the density matrix of a single spin. When the index of the affected \( k \)th spin is arranged along the block-diagonal structure of the overall density matrix \( \hat{\rho} \), the corresponding dissipative contribution couples only the matrix elements lying on the same off-diagonal within each block-diagonal sector. These elements do not contribute to nontrivial many-body correlations and are therefore irrelevant to quantum synchronization. For the second part of the Lindblad super-operator, the anticommutator term \( \{ \hat{\mathcal{A}}^\dagger \hat{\mathcal{A}}, \hat{\rho} \} \) only induces an energy shift in the diagonal elements of \( \hat{\rho} \); that is, it does not couple different elements of \( \hat{\rho} \) and then does not affect the system's coherence relevant to quantum synchronization. Therefore, the nonzero coherence blocks of \( \hat{\rho}^{(1)} \) are determined by the structure of the commutator \( [\hat{V}, \hat{\rho}^{(0)}] \).

We now identify the steady-state components that contribute to quantum synchronization. To this end, we derive the synchronization measure $S_2$ for two-body systems, defined as
\begin{equation} \label{eq:syncmeasure measure}
    S_2(\phi_{12}) = \int_0^{2\pi} d\phi_2 S_2(\phi_{12}+\phi_2, \phi_2)
\end{equation}
where $S_2(\phi_{12}+\phi_2, \phi_2) = \left\langle c^{J}(\phi_{12}+\phi_2) \otimes c^{J}(\phi_2) \right\rangle - \frac{1}{(2\pi)^2}$, and the matrix elements of operator $c^J(\phi)$ is given by
\begin{align}
    c^J_{n,m}(\phi) =& e^{i(n-m)\phi} \frac{2J+1}{4\pi} \int_0^{\pi} d\theta \sin(\theta) d^J_{n,J}(\theta) d^J_{m,J}(\theta)\\
                    =& \frac{e^{i(n-m)\phi} }{2\pi} \frac{\Gamma(1+J+\frac{n+m}{2})\Gamma(1+J-\frac{n+m}{2})}{\sqrt{(J+n)!(J-n)!(J+m)!(J-m)!}}\\
\end{align}
To investigate the dependence of the synchronization measure Eq. (\ref{eq:syncmeasure measure}) on the relative phase $\phi_{12}$, we factorize the operator as 
\begin{equation}
    c^J_{n,m}(\phi)=e^{i(n-m)\phi} c'_{J,n,m}
\end{equation}
where $c'_{J,n,m}$ is real and independent of $\phi$. Substituting this into the definition of $S_2(\phi_{12})$, we obtain
\begin{align}
    S_2(\phi_{12}) =& \int_0^{2\pi} d\phi_2 (\left\langle c^{J}(\phi_{12}+\phi_2) \otimes c^{J}(\phi_2) \right\rangle - \frac{1}{4\pi^2})\\ 
    =& \sum_{n,m} \sum_{q,p} \delta_{n-m+p-q,0} e^{i(n-m)\phi_{12}} c'_{J,n,m} c'_{J,p,q} \rho_{nm,qp} - \frac{1}{2\pi}\\
\end{align}
Here, $\delta_{n-m+p-q}$ ensures that only the coherence terms in the density matrix that satisfy $n-m+p-q=0$ contribute to quantum synchronization. These terms correspond to correlators of the form
\begin{equation}
    \left\langle (\hat{J}_1^+ \hat{J}_2^-)^x + h.c. \right\rangle, \:\:x = 1,2,...,2J
\end{equation}
highlighting the role of phase-coherent processes in generating synchronization.

Taken together, these results show that the synchronization measure \( S_2 \) receives contributions only from coherence terms of the form \( \langle (\hat{J}_1^+ \hat{J}_2^-)^x + \mathrm{h.c.} \rangle \). 
Since the dissipative dynamics—both jump and anticommutator terms—do not generate such many-body coherences, the relevant off-diagonal elements in the first-order correction \( \hat{\rho}^{(1)} \) arise solely from the commutator \( [\hat{V}, \hat{\rho}^{(0)}] \). 
In particular, only interaction terms in \( V \) that explicitly contain \( \hat{J}_1^+ \hat{J}_2^- + \mathrm{h.c.} \) contribute to synchronization. 
Therefore, the emergence and magnitude of \( S_2 \) are entirely governed by these phase-correlated interaction channels, while other components of \( V \) or the dissipative terms remain irrelevant to quantum synchronization.

\begin{figure}[t]
    \centering
    \includegraphics[width=15cm]{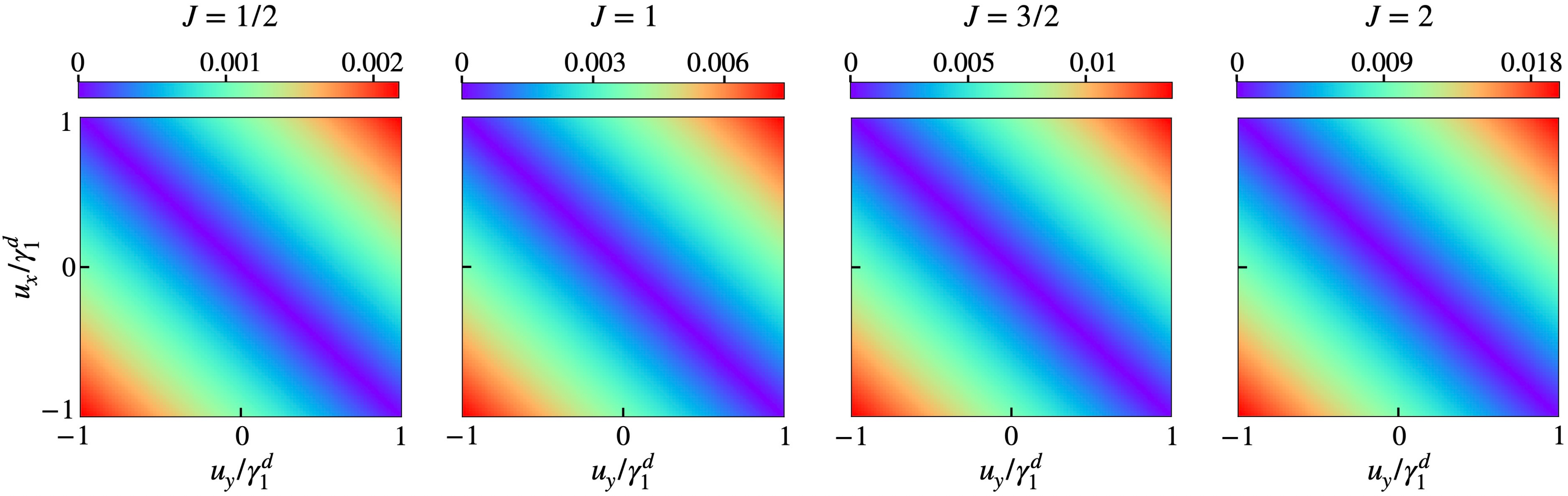}
    \caption{Quantum synchronization measures for two-body systems with different spin values under anisotropic interactions. Parameters: $\epsilon = 0.1\gamma_1^d$, $\gamma_1^g = \gamma_2^d = 2\gamma_1^d = 2\gamma_2^g$.}
    \label{figS1}
\end{figure}

\subsection{Illustrative examples}
In Fig.~S1, we present the quantum synchronization measure \( S_2 \) for two-body systems with spin values \( J = 1/2, 1, 3/2, 2 \) under anisotropic interactions. In this section, the Lindblad operators are defined as
\begin{equation}
\hat{O}^{\pm}_{k} = \hat{J}_{k}^{\pm},
\end{equation}
where \( \hat{J}_{k}^{\pm} = \hat{J}_{k}^{x} \pm i \hat{J}_{k}^{y} \) are the raising and lowering operators for the \( k \)th spin. This choice differs from that in the main text, where the Lindblad operators are given by \( \hat{O}^{\pm}_{k} = \hat{J}_{k}^{\pm} \hat{J}_{k}^z \). The interaction strength is set to \( \epsilon = 0.1\gamma_1^d \), with dissipation rates \( \gamma_1^g = \gamma_2^d = 2\gamma_1^d = 2\gamma_2^g \). The results show that the quantum synchronization blockade persists across different spin values, highlighting the robustness of the phenomenon.

\section{Macroscopic Quantum synchronization and its blockade}
To investigate the impact of anisotropic interactions on quantum synchronization in oscillator networks, we consider a macroscopic quantum network with all-to-all coupling.
Consistent with the main text, we model each oscillator in the network as a spin-1 particle.
In the frame rotating with the common frequency $\omega_z$, The dynamics of the network are governed by the quantum master equation:
\begin{align}\label{eq:mastereq}
\frac{d\hat{\rho}^{(N)}}{dt} &= -i[\frac{\epsilon}{N}\sum_{k<l}(u_x \hat{J}_k^x \hat{J}_l^x + u_y \hat{J}_k^y \hat{J}_l^y), \hat{\rho}^{(N)}] + \sum_{k=1} (\gamma_+ \mathcal{D}\left[\hat{J}_k^+ \hat{J}_k^z\right] + \gamma_- \mathcal{D}\left[\hat{J}_k^- \hat{J}_k^z\right])\hat{\rho}^{(N)}
\end{align}

Due to the exponential scaling of the Hilbert space with system size, direct simulation of the master equation becomes computationally intractable for large $N$. 
To address this, we adopt a mean-field approach, which yields an exact description of the collective dynamics in the thermodynamic limit of $N\to \infty$ under all-to-all coupling. 
In this approach, we assume all oscillators are identical and uncorrelated, i.e., $\hat{\rho}^{(N)} = \hat{\rho}^{\otimes N}$, allowing us to treat the system as a single representative spin with collective properties. 
Under this approximation, the interaction term in Eq.~(\ref{eq:mastereq}) reduces to an effective single-body operator:
\begin{equation}
    \frac{\epsilon}{N} \sum_{k<l} \left( u_x \hat{J}_k^x \hat{J}_l^x + u_y \hat{J}_k^y \hat{J}_l^y \right) \rightarrow \epsilon \left( u_x \hat{J}^x \langle \hat{J}^x \rangle + u_y \hat{J}^y \langle \hat{J}^y \rangle \right)
\end{equation}
Each dissipation term in Eq.~(\ref{eq:mastereq}) acts locally on an individual spin. By tracing out the degrees of freedom of all other spins, $Tr_{(N-1)}[\mathcal{D}[\hat{J}^\pm \hat{J}^z]\hat{\rho}^{\otimes N}] = \mathcal{D}[\hat{J}^\pm \hat{J}^z]\hat{\rho}$, 
the effective contribution to the representative spin is given by
\begin{equation}
\mathcal{D}[\hat{J}_k^\pm \hat{J}_l^z] \hat{\rho}^{(N)} \longrightarrow \mathcal{D}[\hat{J}^\pm \hat{J}^z] \hat{\rho}.
\end{equation}
This replacement holds identically for all sites due to the symmetry of the mean-field ansatz.
Consequently, the resulting dynamics reduces to a nonlinear master equation for a representative spin:
\begin{align} \label{eq:meanfield1}
\frac{d\hat{\rho}}{dt} &= -i[\epsilon(u_x \hat{J}^x \langle \hat{J}^x\rangle + u_y \hat{J}^y \langle \hat{J}^y\rangle), \hat{\rho}] + (\gamma_+ \mathcal{D}\left[\hat{J}^+ \hat{J}^z\right] + \gamma_- \mathcal{D}\left[\hat{J}^- \hat{J}^z\right])\hat{\rho}
\end{align}

To characterize the state of the system, we employ the average amplitude $\langle \hat{J}^+ \rangle = \frac{1}{N} \sum_k \langle \hat{J}_k^+\rangle$, where $\langle \hat{J}_k^+\rangle = e^{i\phi_k}|\langle \hat{J}_k^+\rangle|$. 
To obtain our results, we numerically time-integrate the nonlinear master equation (\ref{eq:meanfield1}). 
In analogy with the classical Kuramoto framework for phase oscillators, collective synchronization within the group emerges only when the coupling strength exceeds a certain threshold, denoted as $\epsilon_c$. 
Throughout the main text, we deliberately choose interaction strengths above the critical threshold $\epsilon_c$ identified for the isotropic case. 
By adjusting $u_x$ and $u_y$, the system tends toward a phase configuration with finite amplitude in the long-time limit. In some regions of parameter space, $\operatorname{Re} \langle \hat{J}^+ \rangle$ exhibits persistent oscillations, while in others it converges to a constant value. 
In regions where $\operatorname{Re} \langle \hat{J}^+ \rangle$ exhibits persistent oscillations, we identify the system as synchronized. In contrast, in regions where $\operatorname{Re} \langle \hat{J}^+ \rangle$ settles to a constant value, we interpret the nonzero order parameter as a result of non-synchronization effects.

\section{Experimental requirements and realizations}
A realistic implementation of our dissipative qubit model with anisotropic XYZ interactions can be achieved using arrays of neutral atoms trapped in optical tweezers. The qubit is encoded in two Zeeman sublevels of the $F=1$ hyperfine manifold, specifically $\{ |-1\rangle, |+1\rangle \}$ of the electronic ground state. These two states are coherently and selectively coupled to distinct Rydberg states via polarization-resolved dressing lasers, adapting the two-color dressing scheme developed by Steinert \textit{et al.}~\cite{PRL2023Arrays}.
Each qubit state $|m\rangle$ is off-resonantly coupled to a Rydberg state $|r_m\rangle$ using laser beams with Rabi frequency $\Omega_m$ and detuning $\Delta_m$, where $m = \{-1, +1\}$. The transitions are addressed via appropriate polarizations ($\sigma^+$, $\sigma^-$), enabling full two-level connectivity. The Rydberg pair states introduce interaction-induced energy shifts, and adiabatic elimination of singly and doubly excited states leads to effective fourth-order processes within the qubit manifold.

This results in an effective interaction Hamiltonian of the form
\begin{equation}
\hat{H}_{\text{int}} = \sum_{k<l} \epsilon \left(
u_{kl}^{+-} \hat{J}^+_k \hat{J}^-_l + u_{kl}^{++} \hat{J}^+_k \hat{J}^+_l
+ \text{H.c.}
\right),
\end{equation}
where $\hat{J}^{\pm}$ are the standard Pauli operators acting on the two-level system. The coefficients $u_{kl}^{+-}$ and $u_{kl}^{++}$ arise from fourth-order virtual processes mediated by laser dressing and van der Waals interactions between Rydberg states. Explicitly~\cite{PRL2023Arrays},
\begin{align}
u_{kl}^{++} &= \sum_{\alpha} \frac{(\Omega_{+} \Omega_{-})^2}{4 \Delta_{+} \Delta_{-}} \cdot \frac{c^{\alpha}_{++} \, c^{\alpha}_{--}}{\Delta^{(2)}_\alpha}, \\
u_{kl}^{+-} &= \sum_{\alpha} \frac{(\Omega_{+} \Omega_{-})^2}{16 (\Delta_{+} \Delta_{-})^2} (\Delta_{+} + \Delta_{-})^2 \cdot \frac{c^{\alpha}_{+-} \, c^{\alpha}_{-+}}{\Delta^{(2)}_\alpha},
\end{align}
where $c^{\alpha}_{mn}$ is the overlap of the asymptotic Rydberg pair state $|r_m r_n\rangle$ with the molecular eigenstate $|\Psi^{(2)}_\alpha\rangle$, and $\Delta^{(2)}_\alpha$ includes the interaction-induced energy shift of the Rydberg pair state. 
These expressions allow the anisotropy ratio $J_{ij}^{++}/J_{ij}^{+-}$ to be tuned by varying the laser detunings $(\Delta_{\pm})$, Rabi frequencies $(\Omega_{\pm})$, and the spatial configuration (angle and distance) of the atom pair.

By varying the laser detunings and Rabi frequencies over typical experimental ranges, e.g., $\Omega_{\pm} = 2\pi \times (0.3 - 0.6)$~MHz and $\Delta_{\pm} = 2\pi \times (1 - 5)$~MHz, both interaction strengths $u_{kl}^{++}$ and $u_{kl}^{+-}$ can be tuned independently from near zero up to several KHz:
\[
u_{kl}^{++}, u_{kl}^{+-} \in 2\pi \times [0, 10]\,\mathrm{kHz}.
\]
In particular, setting $\Delta_+ = -\Delta_-$ leads to destructive interference of flip-flop excitation paths and yields $u_{kl}^{+-} \simeq 0$, corresponding to a fully anisotropic XYZ interaction dominated by $u_{kl}^{++}$. Conversely, by choosing symmetric detunings and appropriate Rydberg pair detunings, one can suppress flop-flop processes and realize $u_{kl}^{++} \simeq 0$, recovering a conventional isotropic XYZ model with only $u_{kl}^{+-}$. The relative weights of $u_{kl}^{++}$ and $u_{kl}^{+-}$ thus define a continuously tunable anisotropy axis in the XYZ plane.

To introduce local dissipation in the two-level system, we engineer optical pumping channels between the two qubit states. The states $|\pm1\rangle$ are weakly driven to a short-lived excited state $|e\rangle$ (e.g., $5P_{3/2}$) using $\sigma_\mp$ polarized lasers, and subsequently decay to the opposite qubit state via spontaneous emission. This produces effective jump operators
\[
\hat{O}^{+}_k = |-1\rangle_k \langle +1|, \quad \hat{O}^{-}_k = |+1\rangle_k \langle -1|,
\]
with corresponding dissipation rates $\gamma_+$ and $\gamma_-$, which are independently tunable through the Rabi frequencies and detunings of the dressing lasers.

This platform enables the exploration of non-equilibrium many-body dynamics in a fully connected qubit network, with tunable coherent anisotropic XYZ interactions and engineered local dissipation. All relevant energy scales—$u_{kl}^{\pm\pm}$ and $\gamma_{\pm}$—are accessible in current experiments, with coherent coupling strengths on the order of $2\pi \times [1$–$10]$~kHz and dissipative rates tunable from tens to hundreds of kHz. The system thus provides a versatile platform for investigating dissipative synchronization blockade and other emergent phenomena.

\begin{figure}[t]
    \centering
    \includegraphics[width=15cm]{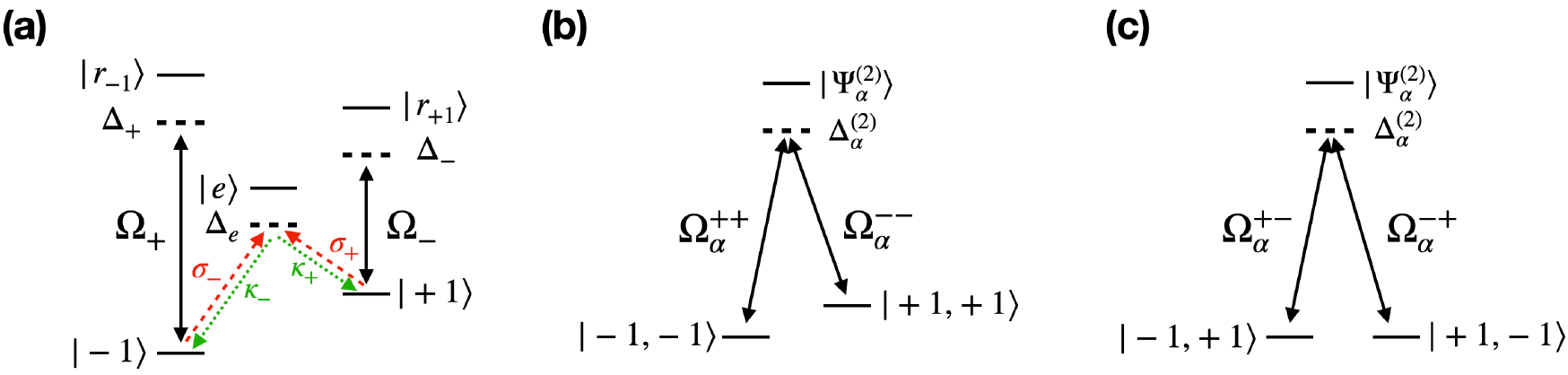}
    \caption{Experimental setup and level schemes. (a) On the single-atom level, we couple the electronic ground states ($|m_F=-1,+1\rangle$) to distinct Rydberg states ($|r_{-1}\rangle, |r_{+1}\rangle$) using polarization-resolved dressing lasers with Rabi frequencies $\Omega_{-}, \Omega_{+}$ and detunings $\Delta_{-}, \Delta_{+}$. Dissipative transitions between \( |m_F = +1\rangle \) and \( |m_F = -1\rangle \) are implemented via optical pumping through short-lived excited states, with tunable rates \( \sigma_+ \) and \( \sigma_- \), respectively.  The excited state \( |e\rangle \) decays spontaneously to the opposite ground state with decay rates \( \kappa_+ \) and \( \kappa_- \). 
    (b) Schematic for the flop-flop interaction \( u_{kl}^{++} \) between two atoms \( k \) and \( l \). Pairs of ground-state atoms in \( |+1,+1\rangle \) or \( |-1,-1\rangle \) are off-resonantly coupled to interacting Rydberg pair states \( |\Psi_\alpha^{(2)}\rangle \) with effective two-body Rabi frequencies \( \Omega_{++}^{(\alpha)} \) and \( \Omega_{--}^{(\alpha)} \). The effective two-photon detuning \( \Delta_\alpha^{(2)} \) to each molecular state \( |\Psi_\alpha^{(2)}\rangle \) includes interaction-induced energy shifts between the Rydberg levels. 
    (c) Schematic for the flip-flop interaction \( u_{kl}^{+-} \) between two atoms \( k \) and \( l \). Ground-state atom pairs in \( |+1,-1\rangle \) or \( |-1,+1\rangle \) are virtually coupled to interacting Rydberg pair states \( |\Psi_\alpha^{(2)}\rangle \) via polarization-resolved laser dressing, with effective two-body Rabi frequencies \( \Omega_{+-}^{(\alpha)} \) and \( \Omega_{-+}^{(\alpha)} \). The associated detuning \( \Delta_\alpha^{(2)} \) includes the interaction-induced energy shift of the molecular state \( |\Psi_\alpha^{(2)}\rangle \).
}
    \label{figS2}
\end{figure}

\end{document}